# MONOLAYER CHARACTERISTICS OF CHITOSAN ASSEMBLED IN LANGMUIR FILMS MIXED WITH ARACHIDIC ACID

Jayasree Nath, R. K. Nath*

*Department of Chemistry, Tripura University, Suryamaninagar – 799022, Tripura, India*
*Email: rknath1958@gmail.com*

Adrita Chakraborty[2]
*Indian association for the Cultivation of Science, Jadavpur, Kolkata, Westbengal*

Syed Arshad Hussain*

*Department of Physics, Tripura University, Suryamaninagar – 799022, Tripura, India*
*Email: sa_h153@hotmail.com, sahussain@tripurauniv.in*

Here we report our investigating on the monolayer characteristics of a non-toxic, biocompatible, and biodegradable polymer chitosan (CHS) using Langmuir – Blodgett technique. It has been observed that pure CHS do not form stable monolayer. However, when CHS is mixed with arachidic acid (AA) stable self supporting monolayer is formed at air-water interface. This also can be transferred onto solid substrate. CHS-AA mixed monolayer is extremely stable with variation of pH. Atomic force microscopy study confirms the formation of stable uniform CHS-AA films onto solid support.

## 1 Introduction

The use of naturally occurring and renewable substances increases day by day in the field of biomedical application in recent years. In that sense chitosan is very important because of its biological properties. Chitosan (CHS) is a non-toxic, biocompatible, and biodegradable polymer[1]. Molecular structure of CHS is shown in the inset of figure 1. CHS and its derivatives have most distinguishable properties; it has extensively been using in the gene delivery[2-5] due to its biodegradability and biocompatibility[1]. Chitosan has good mucoadhesive property that is why it has been used in drug delivery[6] in colonic part of human body. It is usually obtained by deacetylation of the most abundant naturally occurring chitin[7-8], which is extracted from the skeletal structure of crustaceans, insects, mushroom and the cell of fungi. The use of chitin is limited due to its insolubility[9]; where as its derivative chitosan has been extensively used. Chitosan is actually composed predominantly of unbranched chains of β-(1-4)-2- amino-2-deoxy-D-glucopyranose as shown in the inset of figure 1. In case of cell membrane, protein-lipid-protein layer which motivate the use of Langmuir monolayers, deposited Langmuir-Blodgett films[10] can be used to move highly organized lamellar lipid stacking, which can also be applied in biosensors[11], bioreactor[12], electrical and photo devices[13]. Moreover, the action of chitosan as a weight reducer[14] involves lipid-polysaccharides interactions. Chitosan has negligible surface activity[15] as it absorbs in lipid monolayer and changes its organization.

Ultrathin films composed of a single layer or multilayer of biomolecules with highly organized nanostructures onto solid surfaces have been drawing increased attention because of their potential applications in a broad range of fields, such as chemical or biosensors, drug delivery, catalysis, nanoelectronic devices and their important role in the fundamental investigation of complex interactions in natural systems[16-18]. The nanostructures, either two dimensional surface features or three dimensional inner alignments, presenting in the artificial biosystems can yield interesting novel chemical and/or physical properties or provide templates for fabricating new materials[19-20].

Langmuir-Blodgett (LB) technique enables to prepare two-dimensional molecular assemblies (Langmuir monolayer) of lipids, proteins and other biomolecules, which have been extensively used as models to understand the role and the organization of biological membranes, and to acquire knowledge about the molecular recognition process, enzymatic catalysis, cell adhesion and membrane fusion etc [21-22].

LB technology also allows building up of uniform stacks / layers of molecules by transferring a monomolecular film onto a solid support, with an accurate control of the thickness and of the molecular organization, which can be the building blocks of nanobiotechnology[22-23]. Therefore, it is extremely important to study different biomolecules assembled onto Langmuir (that ideally mimic the biological membrane) and LB films.

Thin film properties of chitosan depends on its morphology, which is influenced by molecular weight, degree of N-acetylation, solvent evaporation, and free amine regenerating mechanism. In addition, the solvent used has an influence on the properties of chitosan films[24]. Chitosan coatings have been studied by several researchers to improve the quality and extend the storage life of food products[25-26]. Although the biopolymer CHS has been extensively used in biotechnology and drug delivery, however, the monolayer characteristics of CHS using LB technique has never been reported. Therefore, we felt that the investigations of monolayer characteristics of CHS and information about the stability may be extremely useful for its biological applications especially in drug delivery[6]. Accordingly we have studied the monolayer characteristics of positively charged poly saccharide chitosan (CHS) mixed with a long chain fatty acid arachidic acid (AA).

In this paper we report the monolayer characteristics of CHS mixed with AA at air-water interface and in LB films. It has been observed that CHS-AA mixed films are extremely stable at different pH.

CHS has been exploited as vehicle for drug delivery in various forms, for example, nanoparticles, microparticles, thin films and fibers[27]. CHS tablets as carrier was used to carry prednisolone[28], a water-insoluble drug and Levodopa (used for treatment of perkinson) delivery[29].

On the other hand the choice of biomaterials suitable for forming the carrier film matrix and the barrier films were dictated by various factors, viz., (a) compatibility with the gastric environment, (b) stability during the time of drug delivery, (c) adequate mechanical properties, (d) ease of fabrication and cost, and (e) no appreciable swelling in water and softening point above 37°C[27]. Therefore, our observation about the stability of CHS monolayer at air-water interface may be useful for the biological application of chitosan mainly in drug delivery.

## 2 Materials and method

Deacetylated Chitosan (75 – 85% deaceylated with a low value molecular weight of 22 kd, Molecular structure is shown in the inset of figure 1), Acetic acid (99.5%), Arachidic acid (AA) (99%) were purchased from Sigma Chemicals, USA and used as received. Chloroform (99.9%; SRL, India) was used as a solvent and deionised water (resistivity = 18 MΩ-cm) was used as the sub phase.

The surface area isotherms were recorded in a Langmuir-Blodgett Film deposition Instrument (LB-APEX 2007, India) with symmetrical compression of the monolayers. Freshly prepared samples are always used for measuring the surface pressure – area per molecule isotherm or for film deposition. The trough is equipped with a surface pressure sensor based on the Whilhelmy method. The solution of AA was prepared using chloroform as solvent. Chitosan (CHS) solution was prepared in chloroform with 1% acetic acid. Appropriate amount of either AA or CHS or AA-CHS mixed solution were carefully injected onto the air-water interface of the LB trough with the help of a micro-syringe. After allowing sufficient time to evaporate the barrier was compressed at a speed of 5 mm/min in order to record the pressure – area isotherm or to reach a particular surface pressure for LB film deposition. LB film was deposited onto either cleaned quartz or mica substrate at a surface pressure of15 mN/m.

The atomic force microscopy (AFM) image of CHS-AA mixed film was taken with a commercial AFM system Innova AFM system (Bruker AXS Pte Ltd.) using silicon cantilevers with a sharp, high apex ratio tip (VeecoInstruments). The AFM image presented here was obtained in intermittent-contact ("tapping") mode. The mono- layers on mica substrates were used for the AFM measurement.

## 3 Monolayer characteristics at air-water interface
### 3.1 Concentration dependent study

Pure chitosan (CHS) solution was spread very slowly on the clean water surface of the Langmuir trough and after allowing sufficient time (about 15 min) to evaporate the solvent, the barrier was compressed slowly at a speed of 5 mm/min to study the monolayer characteristics of pure CHS at air-water interface. It is observed that the surface pressure raised only about6.8 mN/m (figure 1). Addition to large amount of CHS solution although improved the isotherm but at the same time led to the formation of microcrystalline domains at the air water interface visible with naked eye. It was also observed that upon relaxation of the surface pressure (by expanding the barrier) this microcrystal broke up into smaller clusters but did not disintegrate completely at the molecular level. Also repeated attempts to transfer this floating layer onto solid support were failed. This suggests that pure CHS did not form stable self supporting monolayer film at air-water interface. However, when AA was mixed with CHS, a stable self supporting monolayer film was formed, which exhibited a pressure-area isotherm with an initial characteristic gaseous region followed by liquid and solid condensed region. Also this CHS-AA mixed film was successfully transferred onto quartz / mica substrate. This observation indicates that CHS when mixed with AA formed stable monolayer at air-water interface. This is because here AA molecules act as a matrix to host the CHS molecules in the mixed films. There may exists some interaction between CHS and AA, which is most likely due to the formation of intermolecular hydrogen bonding between the amino and hydroxyl groups in the CHS and the hydroxyl groups in the fatty acids[30].

Figure 1 shows the representative CHS-AA mixed isotherm at three different mole fraction of CHS viz. 0.5 M, 0.25 M, and 0.125 M along with the pure CHS and AA isotherm for comparison. It has been observed that the pure AA isotherm is consistent with the behavior similar to the previous measurements of this system[31]. The AA isotherm indicates that surface pressure increases rapidly and there exists a transition point at which molecular area is compressed to its limit. This point marks the liquid-solid transition of the molecular layer, at which molecules are densely packed and the surface is no longer compressible. The transition point from liquid to solid state for AA molecules occur at a pressure of about 21.5 mN/m. Isotherm of pure AA gives the limiting area ($A_0$) of 0.22 $nm^2$. The value of $A_0$ was calculated by extrapolating the solid condensed region of the isotherm to the area per molecule axis (X axis) of the surface pressure – area per molecule isotherm. This value is reasonable; since it is well known that long-chain fatty acids occupy the molecular area close to 0.20 $nm^2$ when stand vertically on water surface[32].

mole fraction, (b) shows the plot of collapse pressure as a function of CHS mole fraction. Molecular structure of CHS is also shown in the inset of Fig 1

The mixing of AA at different mole fraction of CHS caused an expansion of the Langmuir monolayer resulting the shifting of the CHS-AA mixed isotherms towards larger area. This effect was prominent for 0.5 M of CHS with the largest expansion causing an appreciable change in the extrapolated area of about 0.09 $nm^2$ (0.31–0.22 $nm^2$, table 1). The interesting thing is that CHS-AA mixed isotherms possess three distinct phases – (i) liquid phase (ii) liquid expanded and (iii) solid condensed phases. In case of pure AA isotherm the liquid expanded phases was not distinguishable. It is interesting to note that for all AA-CHS mixed isotherm there is a plateau like transition point between the surface pressure 23 – 30 mN/m regions. This is also the transition between liquid – expanded to solid phase of the mixed monolayer films. This plateau like transition is an indication of some reorientation of the CHS molecules occurred in the AA-CHS mixed monolayer at this surface pressure when the monolayer films goes to condensed state from the liquid expanded states.

It is interesting to mention in this context that AA is a long chain fatty acid having long hydrophobic chain as a tail and hydrophilic carboxylic acid (-COOH) group as a head group and are ideal material for Langmuir monolayer formation[33]. AA forms stable compact monolayer at air-water interface showing three distinct (gas, liquid and solid phase) phases in the pressure area isotherm[33]. On the other hand CHS is a biopolymer[9]. The AA-CHS mixed isotherm reflects the behavior of CHS along with the AA. After obtaining the liquid phase for the AA-CHS mixed films liquid expanded phase was observed. This is because there may occur some reorientation of CHS molecules in the mixed films after achieving the liquid phase. Such liquid expanded phases were also reported for several other molecules in Langmuir monolayer and explained due to the occurrence of certain reorientation of the constituent molecule in the monolayer[34-35].

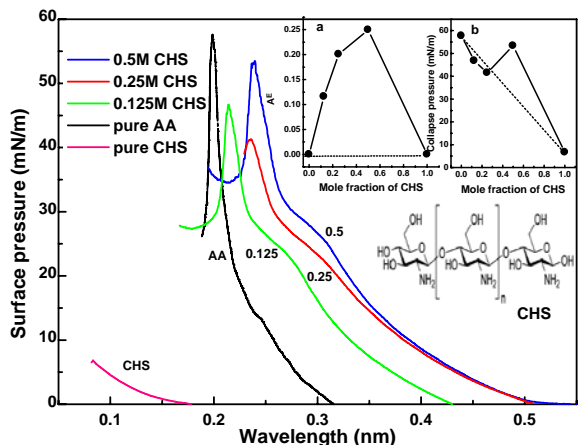

Fig.1 Plot of surface pressure versus area per molecule isotherms of CHS-AA mixed monolayer at different mole fractions of CHS in AA matrix viz. 0.5, 0.25, 0.125 along with the pure AA and pure CHS isotherm. The inset (a) shows the plot of excess area ($A^E$) as a function of CHS

Table 1 Extrapolated Area ($A_{ex}$), Area of the "onset of condensation" ($A_0$) and collapse pressure ($A_C$) for different chitosan concentrations.

| Substance | Mole fraction of CHS | $A_0$ (nm$^2$) | $A^E$ (nm$^2$) | $A_C$ (mN/m) |
| --- | --- | --- | --- | --- |
| AA | 0 | 0.22 | 0 | 57.7 |
| AA+CHS | 0.5 | 0.31 | 0.116 | 46.8 |
| AA+CHS | 0.25 | 0.29 | 0.201 | 41.5 |
| AA+CHS | 0.125 | 0.26 | 0.249 | 53.39 |
| AA+CHS | 1 | - | 0 | 6.7 |

In order to have more information about the AA-CHS mixed monolayer, we have studied the mixing behaviour using additivity and surface phase rule [36-37]. It relevant to mention in this context that the thermodynamic nature of the mixing of various two component systems can be established by analyzing the pressure versus area per molecule isotherm of the pure components as well as the binary mixtures[36]. The excess areas of mixing $A^E$ which provides a measure of non-ideality can be calculated as a function of surface pressure, mixture of composition and the molecular area using the additivity rule[37] as follows:

$$A^E = A_{12} - A_{id}$$
$$A_{id} = A_{CHS} N_{CHS} + A_{AA} N_{AA}$$

Where, $A_{12}$ is the actual area per molecule of mixed monolayer extracted from the mixed isotherm, $A_{id}$ is the calculated average area per molecule of mixture, assuming mole fraction additivity, $A_{CHS}$ and $A_{AA}$ are the areas per molecule (or monomer) of each of the single component i.e. CHS and AA at a specific surface pressure, $N_{CHS}$ and $N_{AA}$ are their corresponding mole fractions in the mixed monolayer films ($A_{CHS} + A_{AA} = 1$).

In the ideal case, the plot of $A_{12}$ versus $A_{CHS}$ will be a straight line. Any deviation from the straight line ($A^E = A_{12} - A_{id} \neq 0$) indicates partial miscibility and non-ideality[38]. If attractive intermolecular forces among the constituent molecules are dominant, $A^E$ will be negative. On the other hand, positive values ($A^E > 0$) indicate a repulsive interaction[31] between the constituent components (i.e. in between CHS and AA) of the mixed monolayer.

Inset a of figure 1 shows the plot of excess area ($A^E$) as a function of CHS mole fraction ($N_{CHS}$) in case of CHS-AA mixed monolayer films at a surface pressure of 5 mN/m. From inset a of figure 1, a noticeable deviation from the additivity rule is observed, which is an indication of interaction among the constituent molecules of the binary mixture in the mixed monolayer. The observed deviation is predominantly positive, meaning that repulsive interactions are dominant among the CHS and AA molecules in the CHS-AA monolayer films. The extent of interaction increases with the increase in mole fraction of CHS. These repulsive interactions among the constituent molecules are mainly responsible for the aggregation in the mixed films even at very low concentration of CHS. This is consistent with the behaviour of other non-amphiphiles[39].

According to the surface phase role relationship between collapse pressure ($\pi_C$) with mole fraction give the information about 2D miscibility in the binary mixed monolayer[36]. The plot of $\pi_C$ as a function of CHS mole fraction is also shown in the inset b of figure 1,where the dashed line represent the ideality characteristics indicating the miscibility of the constituent molecules. From the figure it has been observed that at higher mole fraction de-mixing occurred between CHS and AA in the mixed monolayer. Although at lower mole fraction there exists some extent of miscibility in the mixed monolayer. This de-mixing leads to the formation of nanoscale aggregates in the CHS-AA mixed films. Our later atomic force microscopy studies also confirm this.

### 3.2 Hysteresis study

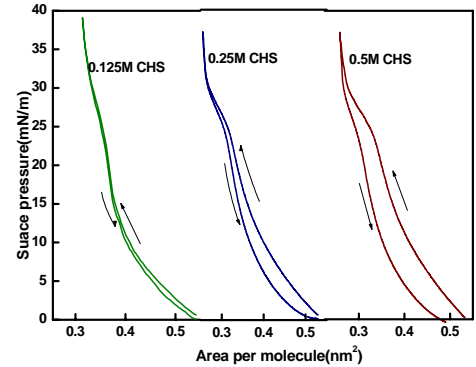

Fig.2 Compression- decompression cycle of CHS-AA mixed langmuir monolayer on the air water interface at different mole fraction of CHS in the CHS-AA mixed monolayer

In order to have more idea about the monolayer stability for the CHS-AA mixed monolayer at air-water interface we have investigated the hysteresis behaviour of the mixed monolayer. This has been done by compressing the mixed monolayer up to a particular surface pressure followed by the expansion of the monolayer. Figure 2 shows the isotherms during compression-expansion cycle for the mixed systems under investigation at three mole fractions. From figure 2 it has been observed that the compression-expansion cycle of the film formed with lower mole fraction of CHS is very narrow, with no significant hysteresis indicates that the film is stable and reversible under these conditions. But with the increase in CHS mole fraction the hysteresis between the compression-expansion curves increase and become maximum at the CHS mole fraction of 0.5 M. This indicates an incomplete reversibility of the film. On the other hand the AA monolayer film shows no hysteresis [figure not shown] with overlapping compression – expansion isotherm. This is expected for long chain fatty acids, as they form stable monolayer films[36]. On the other hand, at the higher mole fraction in the mixed film the number of CHS moiety increases. Again CHS molecules show a change in orientation in monolayer with the surface pressure (as evident from the isotherm of figure 1). Therefore, during the expansion, reorientation of the CHS molecules may take place. The extent of reorientation is large for the higher mole fraction of CHS causing certain amount of hysteresis.

### *3.3 Effect of pH on CHS–AA mixed monolayer*

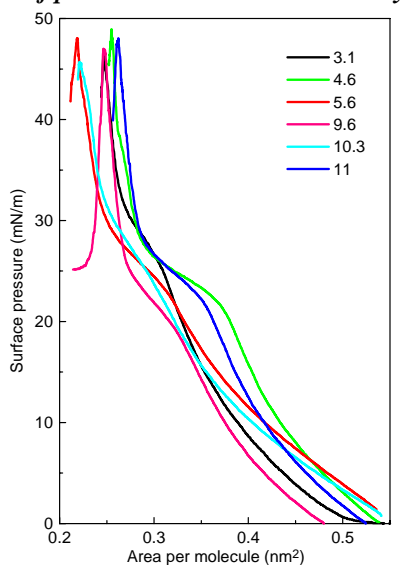

Fig.3 Plot of surface pressure versus area per molecule isotherms of CHS-AA mixed monolayer at da mole fractions of 0.25 M of CHS in AA measured at different pH viz. 3.1, 4.6, 5.6, 9.6, 10.3 and 11

The biopolymer CHS used in the present study has been extensively used in biotechnology and drug delivery[40]. Therefore, it is very important to see the behavior of CHS in the monolayer at different pH. Accordingly we have investigated CHS-AA mixed films at different pH. Our investigations show that mixed films of all the mole fractions (0.125, 0.25 and 0.5 M of CHS) of the mixed films are stable with variations of pH. Figure 3 shows representative images of CHS-AA mixed isotherms for 0.25 mole fraction of CHS measured at different pH viz. 3.1, 4.6, 5.6, 9.6, 10.3 and 11. From figure 3, it has been observed that there is certain variation of areas of the isotherm depending on the pH. However, the nature and shape of the isotherm remains almost same. Interestingly for all the isotherms of 0.25 M, the collapse pressure is increased. As a whole, from these investigations we confirm that CHS-AA monolayer is stable at different pH.

It is relevant to mention that Amino group of CHS has pKa value 6.5. This leads to the protonation of CHS in acidic to neutral solution with a charge density depending on pH[41]. Accordingly, the solubility of CHS changes with pH. At lower pH the solubility of CHS in water increases[9]. At normal pH (6.4-7) CHS remains as zwitterions. With decrease in pH, the degree of ionization increases. This results an increase in solubility of CHS in water. However, at higher pH (4.0 to higher), the degree of ionization of CHS decreases, which makes CHS insoluble[42] in water. As a result the interaction between CHS and AA changes with pH. Accordingly the number of molecules on the air-water interface changes, resulting a change in area per molecule in the floating monolayer.

## 4 AFM characterizations

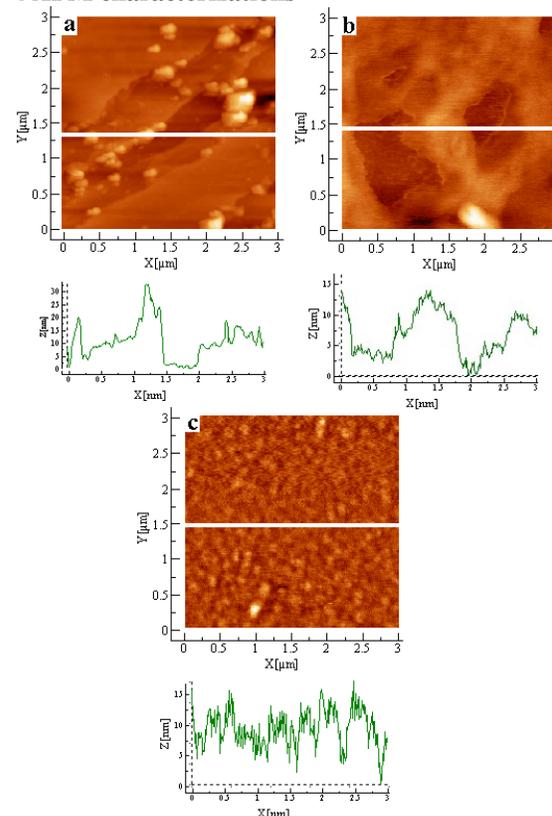

Fig.4 AFM image of (a) pure CHS (deposition pressure 5 mN/m) (b) CHS-AA mixed (0.5 M) films deposited at 5

mN/m and (c) CHS-AA mixed (0.5 M) films deposited at 30 mN/m. Height profile along the line shown in the images are also given

In order to have an idea about the morphology of the surface of the chitosan film transferred onto solid substrate, we have measured the AFM images of the monolayer films. Figure 4 shows the AFM images along with the roughness analysis of pure CHS and CHS-AA mixed films. The pure CHS film has deposited at 5 mN/m surface pressure onto smooth silicon wafer surface for AFM measurement. The AFM image of pure CHS film (figure 4a) shows very rough surface with microcrystalline aggregates having height within 30 nm range. Our previous isotherm studies also supported that pure CHS did not form stable self supporting monolayer at air-water interface. However, it has been seen that when AA is mixed with CHS, self supporting monolayer is formed. The AFM image (figure 4a) confirmed that pure CHS form higher order large size aggregate and multilayer in ultrathin films. Figure 4b shows the AFM image of CHS-AA mixed films (0.5 M) deposited at 5 mN/m surface pressure. This image suggests that the film surface improves significantly. The film thickness is within 13 nm range. This height includes the height of CHS plus height of AA in the mixed films. However, the surface coverage is not very good. In few cases ararchidic acid background is clearly visible. Interestingly the surface coverage is significantly improved when the CHS-AA (0.5 M) mixed film is transferred onto solid surface at a surface pressure of 30 mN/m (figure 4c). The film surface is very uniform and coverage is also very good.

**Conclusion**

In this work, we have investigated the interaction of arachidic acid with chitosan at the air water interface. The result shows pure CHS alone do not form self supporting film at air-water interface. However, when CHS is mixed with a fatty acid AA it form self supporting stable monolayer at air-water interface which can be transferred onto solid substrate. There exists repulsive interaction between CHS and AA in the mixed monolayer leading to the formation of aggregates. Hysteresis studies confirmed the stability of the mixed film at air-water interface. The significant thing is that, CHS – AA mixed films are very stable with variation of pH. AFM measurement of the LB monolayer of pure CHS and CHS-AA mixed monolayer give compelling visual evidence that inclusion of AA improves the CHS film formation. The potential of CHS for the use of carriers in drug delivery has already proved[27]. Our result may be useful for the biological application of chitosan mainly in drug delivery. Since, the stability during the time of drug delivery is one of the important criteria for the choice of biomaterials suitable for forming the carrier film matrix and the barrier films[27].

**Acknowledgements**

The author R. K. Nath acknowledges the financial support from CSIR, Govt. of India through CSIR project No. 01(2289)/08/EMR-II dt. 20-11-2008.